\documentstyle[preprint,eqsecnum,aps,prd,tighten]{revtex}

\input boxedeps

\SetepsfEPSFSpecial
\HideDisplacementBoxes

\newcommand{\half}{{\scriptstyle \frac{1}{2}}}
\newcommand{\beq}{\begin{equation}}
\newcommand{\eeq}{\end{equation}}
\newcommand{\bea}{\begin{eqnarray}}
\newcommand{\eea}{\end{eqnarray}}
\newcommand{\etal}{{\em et al.}}
\newcommand{\ie}{{\em i.e.}}
\newcommand{\etc}{{\em etc.}}
\newcommand{\eg}{{\em e.g.}}

\newcommand{\gevcc}{\hbox{ GeV}\!/\!c^2}
\newcommand{\gev}{\hbox{ GeV}}
\newcommand{\ev}{\hbox{ eV}}

\newcommand{\tev}{\hbox{ TeV}}
\newcommand{\pev}{\hbox{ PeV}}
\newcommand{\cm}{\hbox{ cm}}
\newcommand{\km}{\hbox{ km}}
\newcommand{\pb}{\hbox{ pb}}

\newcommand{\kmwe}{\hbox{ kmwe}}

\newcommand{\eqn}[1]{(\ref{#1})}

\def\slashiii#1{\setbox0=\hbox{$#1$}#1\hskip-\wd0\hbox to\wd0{\hss\sl/\/\hss}}
\newcommand{\rpv}{$\slashiii{R}$}

\newcommand{\rpar}{$R$-parity}

\def\bentarrow{\:\raisebox{1.1ex}{\rlap{$\vert$}}\!\rightarrow}

\def\dk#1#2#3{
        \begin{equation}
        \begin{array}{r c l}
        #1 & \rightarrow & #2 \\
         & & \bentarrow #3
        \end{array}
        \end{equation}
                }
\def\dkdk#1#2#3#4{
        \begin{equation}
        \begin{array}{r c l}
        #1 & \rightarrow & #2 \\
         & & \bentarrow #3 \\
         & & \phantom{\;\bentarrow}\bentarrow #4
        \end{array}
        \end{equation}
                }
\def\dkp#1#2#3#4{
        \begin{equation}
        \begin{array}{r c l}
        #1 & \rightarrow & #2#3 \\
         & & \phantom{\; #2}\bentarrow #4
        \end{array}
        \end{equation}
                }

\renewcommand{\pl}[3]{Phys. Lett.\  {\bf #1,} #2 (19#3)}        
\renewcommand{\prl}[3]{Phys. Rev. Lett.\ {\bf #1,} #2 (19#3)}

\newcommand{\pr}[3]{Phys. Rev. D{\bf #1,} #2 (19#3)}
\newcommand{\np}[3]{Nucl. Phys.\ {\bf #1,} #2 (19#3)}
\newcommand{\npbps}[3]{Nucl. Phys.	B (Proc. Supp.) {\bf #1,} #2 (19#3)}	 %
\newcommand{\zp}[3]{Z.~Phys.\ {\bf C#1,} #2 (19#3)}
\newcommand{\astropp}[3]{Astropart. Phys.\ {\bf #1,} #2 (19#3)}
\newcommand{\ib}[3]{{\em ibid.\/} {\bf #1,} #2 (19#3)}
\newcommand{\philt}[3]{Phil. Trans. Roy. Soc. London A {\bf #1,} #2 (19#3)}							%
\newcommand{\hepph}[1]{(electronic archive: hep--ph/#1)}
\newcommand{\hepex}[1]{(electronic archive: hep--ex/#1)}

\def\be{\begin{equation}}
\def\ee{\end{equation}}
\def\barr{\begin{array}}
\def\earr{\end{array}}
\def\dis{\displaystyle}

\begin{document}
\preprint{FERMILAB--PUB--98/088--T \quad MRI-PHY/980341
\quad CERN--TH/98--97 \quad	\phantom{CERN}}

\title{Manifestations of $R$-Parity Violation \\
in Ultrahigh-Energy Neutrino Interactions}

\author{Marcela Carena\thanks{E-mail: \textsf{carena@fnal.gov}}}
\address{Theoretical Physics Department, 
Fermi National Accelerator 
Laboratory \\ P.O.\ Box 500, Batavia, Illinois 60510 USA}
\author{Debajyoti Choudhury\thanks{E-mail: 
\textsf{debchou@mri.ernet.in, debchou@mail.cern.ch}}} 
\address{Mehta Research Institute of  Mathematics and Mathematical 
Physics, \\  Chhatnag Road, Jhusi, Allahabad -- 221019, India} 
\author{Smaragda Lola\thanks{E-mail: \textsf{magda@mail.cern.ch}}}
\address{Theory Division,	CERN, CH-1211 Geneva 23, Switzerland}
\author{Chris Quigg\thanks{E-mail: \textsf{quigg@fnal.gov}}}
\address{Theoretical Physics Department, 
Fermi National Accelerator 
Laboratory \\ P.O.\ Box 500, Batavia, Illinois 60510 USA}
\date{\today}
\maketitle

\begin{abstract}
Supersymmetric couplings that do not respect \rpar\ can induce 
significant changes in the 
interaction rates of ultrahigh-energy neutrinos through the 
direct-channel production of superpartner resonances, and can provide new 
sources 
of extremely energetic $\tau$-leptons.  We analyze the possible 
observable consequences of \rpv\ transitions in large-volume neutrino 
telescopes.
\end{abstract}

\narrowtext
\section{Introduction}
By observing long-range muons produced in the charged-current 
reactions $(\nu_{\mu},\bar{\nu}_{\mu})N \rightarrow \mu^{\mp} + \hbox{anything}$, we can hope to 
detect ultrahigh-energy (UHE) neutrinos from extraterrestrial 
sources.  The diffuse spectrum of neutrinos produced in the 
interactions of energetic protons in active galactic nuclei (AGNs), neutrinos 
emitted by gamma-ray bursters, and neutrinos produced as a result of 
pion photoproduction on the cosmic microwave background are among the 
targets of UHE neutrino astronomy \cite{marseille}.
Extensive predictions exist for the charged-current and 
neutral-current cross sections at energies up to $10^{21}\ev$ 
\cite{GQRS,GQRS97,others}, and neutrino telescopes with instrumented 
volumes approaching $1\km^{3}$ are under active study 
\cite{marseille,detectors}.

The primary motivation for neutrino observatories is 
to search for sources and to probe
the processes that produce UHE $\gamma$ rays in AGNs.  
However, unconventional processes can modify the expected interaction rates 
of UHE neutrinos and provide evidence for new particles or new 
interactions of direct interest to particle physics.

In this paper, we explore supersymmetric
processes mediated by \rpar--violating (\rpv) interactions and investigate 
their consequences for UHE neutrino interaction rates.  We find that 
direct-channel production of squarks through \rpv\ 
couplings with ordinary particles can significantly enhance the UHE 
neutrino--nucleon cross sections, and can alter the 
neutral-current to charged-current ratio for neutrino-nucleon 
interactions.  Similarly, direct-channel production of sleptons can 
enhance the neutrino--electron cross section at resonance and may 
provide a new source of UHE $\tau$ leptons.  These effects may, in time, 
be observable in neutrino 
telescopes, and our analysis raises some issues to be borne in mind 
as km$^{3}$-class neutrino observatories come into being.

\section{$R$-parity in supersymmetric theories}
In the standard electroweak theory, gauge invariance suffices to 
forbid terms in the Lagrangian that change either baryon number or 
lepton number.  The most general supersymmetric (SUSY) extension of the 
standard model \cite{why}, in contrast, allows couplings that change lepton 
number or baryon number.  In general, such terms may lead to 
an unacceptably short proton lifetime.  A simple, though \textit{ad hoc,} 
solution to this problem is to impose a discrete symmetry called 
\rpar, which implies a conserved multiplicative quantum number, 
$R\equiv (-1)^{3B+L+S}$, where $B$ is baryon number, $L$ is lepton 
number, and $S$ is spin \cite{ff}.  All ordinary particles are 
\rpar--even, while all superpartners are \rpar--odd.  If \rpar\ is 
conserved, superpartners must be produced in pairs and the lightest 
superpartner, or LSP, is absolutely stable.
  
%
However, we have no principle that requires us to impose \rpar\ on the 
SUSY Lagrangian.  What is more, \rpv\ interactions can improve the 
agreement between theory and precision electroweak measurements 
(\eg, $Z^{0} \rightarrow b\bar{b}, c\bar{c}$), and also offer ready 
explanations \cite{flocons} for experimental anomalies such as the 
high-$Q^{2}$ excess reported at HERA \cite{H1,ZEUS}.  Accordingly, it 
is of interest to consider an \rpar--violating extension of the 
minimal supersymmetric standard model (MSSM).  The most general \rpv\ 
terms in the superpotential consistent with Lorentz invariance, gauge 
symmetry, and supersymmetry
are\footnote{We suppress here the SU(2)$_{\mathrm{L}}$ and SU(3)$_c$ 
indices.  Symmetry under SU(2)$_{\mathrm{L}}$ implies that the first 
term is antisymmetric under $i\leftrightarrow j$, while SU(3)$_c$ 
symmetry dictates that the third term is antisymmetric under 
$j\leftrightarrow k$.  We neglect bilinear terms that mix lepton and 
Higgs superfields \protect\cite{hallsuz}.  Discussions of the 
phenomenological implications of such terms can be found in the 
literature~\protect\cite{BILIN}.  }
\begin{equation}
W_{\not{R}}  =  \lambda_{ijk} L^i L^j \bar{E}^k 
+\lambda^{\prime}_{ijk} L^i Q^j \bar{D}^k 
   + \lambda^{''}_{ijk} \bar{U}^i \bar{D}^j \bar{D}^k 
\label{eq:superpot}
\end{equation}
where $i,j,k$  are generation indices,
$L^i \ni (\nu^{i},e^{\,i})_{\mathrm{L}}$ and  
$Q^i \ni (u^{i},d^{\,i})_{\mathrm{L}}$ are the left-chiral
superfields, and 
$E^i \ni e^{\,i}_{\mathrm{R}}$, $D^i \ni d^{\,i}_{\mathrm{R}}$, and 
$U^i \ni u^{i}_{\mathrm{R}}$ 
are the right-chiral superfields, respectively.
 The Yukawa couplings 
$\lambda_{ijk}$, $\lambda^{\prime}_{ijk}$, and $\lambda^{''}_{ijk}$ 
are \textit{a priori} arbitrary, so the \rpv\ 
superpotential \eqn{eq:superpot} introduces 45 free 
parameters.

The $LLE$ and $LQD$ terms change lepton number, whereas the $UDD$ 
term changes baryon number.  Since our interest here is in UHE 
neutrino interactions, we shall explicitly 
forbid the $UDD$ interactions \cite{leptopar} as the most economical way to 
avoid unacceptably rapid proton decay.  Expanding the superfield 
components in \eqn{eq:superpot}, we obtain the interaction Lagrangians
\begin{equation}
{\mathcal{L}}_{LLE} = \lambda_{ijk} 
\left\{ \tilde{\nu}_{\mathrm{L}}^i e_{\mathrm{L}}^j \bar{e}^k_{\mathrm{R}} + 
        \tilde{e}_{\mathrm{L}}^i \nu_{\mathrm{L}}^j 
        \bar{e}^k_{\mathrm{R}} + \tilde{e}_{\mathrm{R}}^{kc} \nu_{\mathrm{L}}^i e^
j_{\mathrm{L}}  
                  \right\} + \mathrm{h.c.} 
\label{eq:LLE}
\end{equation}
and 
\begin{eqnarray}
{\mathcal{L}}_{LQD} & = & \lambda^{\prime}_{ijk} \left\{
 \tilde{\nu}_{\mathrm{L}}^i d_{\mathrm{L}}^j \bar{d}^k_{\mathrm{R}} -
  \tilde{e}_{\mathrm{L}}^i u_{\mathrm{L}}^j \bar{d}^k_{\mathrm{R}}  
  + \tilde{d}_{\mathrm{L}}^j \nu_{\mathrm{L}}^i 
  \bar{d}^k_{\mathrm{R}} \right. \nonumber \\ & &  \left. -
     \tilde{u}_{\mathrm{L}}^j e_{\mathrm{L}}^i \bar{d}^k_{\mathrm{R}}  
  + \tilde{d}_{\mathrm{R}}^{kc} \nu_{\mathrm{L}}^i d^j_{\mathrm{L}} -  
  \tilde{d}_{\mathrm{R}}^{kc} e_{\mathrm{L}}^i u^j_{\mathrm{L}}  \right\} + 
  \mathrm{h.c.}
\label{eq:LQD}
\end{eqnarray}
The \rpv\ couplings in \eqn{eq:LLE} and \eqn{eq:LQD} modify 
supersymmetric 
phenomenology in several important ways: processes that change lepton 
number are allowed, superpartners can be produced singly, and the 
LSP---now unstable against decay into ordinary particles---is no 
longer constrained to be a neutral color singlet to avoid cosmological 
embarrassments.

The remarkable agreement between present data and standard-model 
expectations implies very restrictive bounds on the strength of \rpv\ 
operators \cite{BGH,gautam,herbi,goity}.  $LQD$-type interactions 
of electron neutrinos or antineutrinos with the first-generation quarks 
found in nucleon targets are highly constrained.  We consider for 
definiteness the case of a 200-GeV$\!/\!c^{2}$ sfermion.  The absence 
of a signal for neutrinoless double-beta decay implies that 
$\lambda^{\prime}_{111} \leq 0.002$, for squark and gluino masses of 
$200\gevcc$.  Universality of the charged-current interactions implies 
that $\lambda^{\prime}_{112,113} \leq 0.04$ (at the $2\sigma$ level).  
The couplings $\lambda^{\prime}_{121,131}$ are also constrained: if 
the Cabibbo mixing in the light-quark sector arises solely from the 
\textit{down} quarks, one may argue that the upper bound on the 
branching ratio for the decay $K \rightarrow \pi \nu \bar{\nu}$ 
\cite{e787} limits $\lambda^{\prime}_{121} < 0.02$ \cite{nunubar}.  If 
on the other hand this mixing arises from the \textit{up} sector, an 
effect would have been seen in neutrinoless double beta decay.  In 
either case, $\lambda^{\prime}_{121}$ cannot be very large.  Finally, 
atomic parity violation gives the constraint $\lambda^{\prime}_{121, 
131} \leq 0.07$.  In view of these constraints, we shall not consider 
the effects of \rpar\ violation on $\nu_{e}N$ and $\bar{\nu}_{e}N$ 
cross sections.

Experimental limits on $LQD$ couplings that involve $\nu_{\mu}$ are 
less restrictive.  So too are the limits on $LLE$ couplings, both for 
$\nu_{\mu}$ and $\nu_{e}$.  To 
illustrate some possible consequences of \rpv\ couplings for UHE 
neutrino interactions, we shall analyze separately the effects of 
$LQD$ terms on $(\nu_{\mu}, \bar{\nu}_{\mu}) N$ scattering and of 
$LLE$ terms on $(\nu_{e,\mu},\bar{\nu}_{e,\mu})e$ scattering \cite{nuLQ}, 
considering in turn the influence of massive squarks and sleptons in 
light of existing constraints on $\lambda^{\prime}_{ijk}$ and 
$\lambda_{ijk}$.  In Table \ref{tab:lambdas} we summarize the 
constraints on the \rpv\ Yukawa couplings that are relevant for our 
studies.  In each example we consider, we shall assume that 
only one \rpv\ coupling can be sizeable \cite{products}.

\section{Neutrino-nucleon scattering}
The dominant mechanisms for producing UHE
photons and neutrinos are expected to be
\dk{p\:(p/\gamma)}{\pi^{0}+ \hbox{anything}}{\gamma\gamma}
and
\dkdk{p\:(p/\gamma)}{\pi^{\pm}+ \hbox{%
anything}}{\mu\nu_{\mu}}{e\nu_{e}\nu_{\mu}\;.}
If $\pi^{+}$, $\pi^{-}$, and $\pi^{0}$ are produced in equal numbers, 
the relative populations of the neutral particles will be 
$2\gamma:2\nu_{\mu}:2\bar{\nu}_{\mu}:1\nu_{e}:1\bar{\nu}_{e}$. There 
are no significant conventional sources of $\nu_{\tau}$ 
and $\bar{\nu}_{\tau}$ \cite{nutau}. 

At low neutrino energies ($E_{\nu} \ll M_{W}^{2}/2M$, where $M_{W}$ is 
the intermediate-boson mass and $M$ is the nucleon mass), differential 
and total cross sections for the reactions $\nu N \rightarrow 
\mu+\hbox{anything}$ are approximately proportional to the neutrino 
energy.  For neutrino energies in the range $10\gev 
\le E_{\nu} \le 1\tev$, the cross sections computed \cite{GQRS} using 
the CTEQ3 parton distributions \cite{CTEQ} are reproduced by simple 
power-law forms: 
\begin{eqnarray}
	\sigma_{\mathrm{CC}}(\nu N) & = & 8.66 \times 
	10^{-39}\cm^{2}\left(\frac{E_{\nu}}{1\gev}\right)^{0.953}
	\nonumber  \\
	\sigma_{\mathrm{NC}}(\nu N) & = & 2.66 \times 
	10^{-39}\cm^{2}\left(\frac{E_{\nu}}{1\gev}\right)^{0.958}
	\nonumber  \\
	\sigma_{\mathrm{CC}}(\bar{\nu}N) & = & 3.83 \times 
	10^{-39}\cm^{2}\left(\frac{E_{\nu}}{1\gev}\right)^{0.982} \nonumber \\
	\sigma_{\mathrm{NC}}(\bar{\nu}N) & = & 1.34 \times 
	10^{-39}\cm^{2}\left(\frac{E_{\nu}}{1\gev}\right)^{0.983}\;.
	\label{leparsig}
	\end{eqnarray}
Above $E_{\nu}\approx 10^{12}\ev$, the gauge-boson propagator 
restricts the momentum transfer $Q^{2}$ to values near $M_{W}^{2}$ and 
damps the cross section.  (Parallel remarks apply to the 
neutral-current reactions $\nu N \rightarrow \nu+\hbox{anything}$ 
mediated by $Z$ exchange.) For neutrino energies in the range 
$10^{15}\ev \le E_{\nu} \le 10^{21}\ev$, good representations of the 
cross sections are given by \cite{GQRS}:
\begin{eqnarray}
	\sigma_{\mathrm{CC}}(\nu N) & = & 2.69 \times 
	10^{-36}\cm^{2}\left(\frac{E_{\nu}}{1\gev}\right)^{0.402}
	\nonumber  \\
	\sigma_{\mathrm{NC}}(\nu N) & = & 1.06 \times 
	10^{-36}\cm^{2}\left(\frac{E_{\nu}}{1\gev}\right)^{0.408}
	\nonumber  \\
	\sigma_{\mathrm{CC}}(\bar{\nu}N) & = & 2.53 \times 
	10^{-36}\cm^{2}\left(\frac{E_{\nu}}{1\gev}\right)^{0.404} \nonumber \\
	\sigma_{\mathrm{NC}}(\bar{\nu}N) & = & 0.98 \times 
	10^{-36}\cm^{2}\left(\frac{E_{\nu}}{1\gev}\right)^{0.410}\;. 
	\label{parsig}  
\end{eqnarray}

The constraints we have reviewed on \rpv\ couplings involving muon 
neutrinos or antineutrinos and up or down quarks permit significant 
modifications to the standard-model cross sections for $\nu_{\mu}N$ 
and $\bar{\nu}_{\mu}N$ interactions.\footnote{Since the cases we 
investigate do not involve first-generation leptons, they have no 
observable consequences for the $e^{\pm}p$ reactions studied at HERA.} 
We choose to limit our considerations to interactions involving 
first-generation quarks, which are the predominant constituents of the 
nucleon.  A similar analysis could be carried out for the minority 
constituents $s, c$, \etc\ The relevant \rpar--violating couplings 
then are $\lambda^{\prime}_{21j}$, which drives reactions mediated 
by right-handed squarks, and $\lambda^{\prime}_{2j1}$, which drives 
reactions mediated by left-handed squarks.

\textit{$\nu_{\mu}N$ interactions.} The charged-current reaction 
$\nu_{\mu}N \rightarrow \mu^{-}+\hbox{anything}$ can receive 
contributions from the $s$-channel formation process 
$\nu_{\mu}d_{\mathrm{L}}\rightarrow \tilde{d}^{k}_{\mathrm{R}} 
\rightarrow \mu^{-}_{\mathrm{L}}u _{\mathrm{L}}$, which involves 
valence quarks, and from $u$-channel exchange of 
$\tilde{d}^{k}_{\mathrm{R}}$ in the reaction $\nu_{\mu}\bar{u} 
\rightarrow \bar{d}\mu^{-}$, which involves only sea quarks.  The 
cross section formulas are collected in the Appendix.  We consider the 
influence of a right-handed squark $\tilde{d}^{k}_{\mathrm{R}}$ with 
mass $\widetilde{m}^{k}_{\mathrm{R}} = 200 (400) \gevcc$ and \rpv\ 
Yukawa couplings $\lambda^{\prime}_{21k}= 0.2 (0.4)$, which are about 
the maximum values allowed.  With these choices for masses and Yukawa 
couplings, the total width of the right-handed squark is 
$\Gamma(\tilde{d}^{k}_{\mathrm{R}}) = 0.32\gev$ for 
$\lambda^{\prime}_{21k} = 0.2$, and 
$\Gamma(\tilde{d}^{k}_{\mathrm{R}}) = 2.55\gev$ for $\lambda 
^{\prime}_{21k} = 0.4$, with equal branching fractions into $\mu^{-}u$ 
and $\nu_{\mu}d$.  We have made here the implicit assumption that the 
squark decays to gauginos are suppressed compared to the 
\rpar--violating decays.\footnote{This is the case for relatively 
large $\mu$ and $M_2$, which result in large gaugino masses.  While 
the Tevatron bounds on second generation scalar leptoquarks are 
impressively tight for light sfermions, they are greatly 
relaxed for smaller branching ratios to leptoquark-type decays and for 
higher sfermion masses.  See Ref.\
\cite{LQtev}.} We will revisit this 
assumption when we discuss the contribution of $LL\bar{E}$ operators to 
neutrino-electron scattering.  Left-handed squarks couple either to 
neutrinos or to charged leptons, but not both, so cannot contribute 
to the charged-current reaction.

\begin{figure}              
        \centerline{\BoxedEPSF{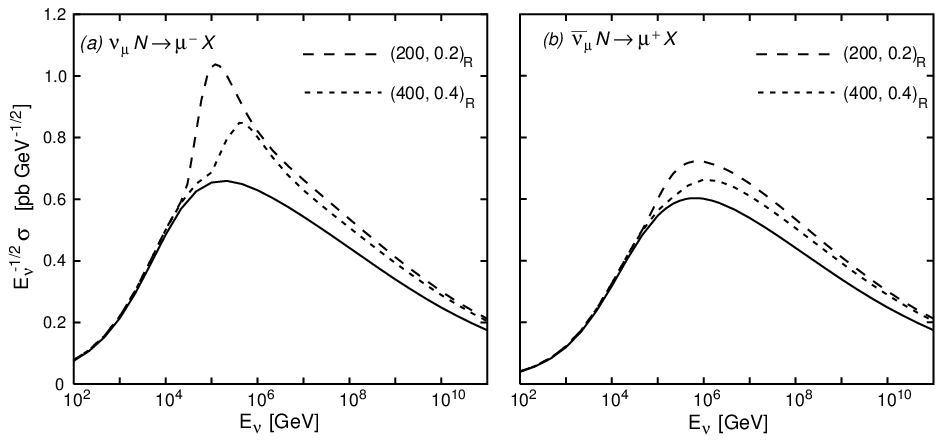 scaled 1350}} 
        \caption[cc]{Charged-current cross sections for (a) 
        $\nu_{\mu}N$ and (b) $\bar{\nu}_{\mu}N$ interactions.  The 
        solid lines show the predictions of the standard model.  The dashed 
        (short-dashed) curves include the contributions of a right-handed 
        squark, $\tilde{d}^{k}_{\mathrm{R}}$, with mass 
        $\widetilde{m}=200\:(400)\gevcc$ and \rpv\ coupling 
        $\lambda^{\prime}_{21k}= 0.2\:(0.4)$.} \label{fig:CC}
\end{figure}

The resonance condition for forming $\tilde{d}^{k}_{\mathrm{R}}$ in 
$\nu q$ interactions is $x = 
\widetilde{m}^{k\;2}_{\mathrm{R}}/2ME_{\nu}$, where $x$ is the 
momentum fraction carried by the quark and $M$ is the nucleon mass.
As a  consequence of the spread in quark momenta, the resonance peaks 
are not narrow, but are 
broadened and shifted above the threshold energies
$E_{\nu}^{\mathrm{th}} = \widetilde{m}^{k\;2}_{\mathrm{R}}/2M \approx 
2.13 \hbox{ and }8.52\times 10^{4}\gev$, respectively.  We have 
calculated the cross sections with and without the \rpv\ contribution 
using the CTEQ3 parton distributions \cite{CTEQ}.  Figure 
\ref{fig:CC}(a) shows that the modifications to the 
standard-model cross section are appreciable: in the two examples we 
consider, the charged-current cross section is enhanced by up to 60\% 
and 30\%, respectively.  Far above the resonance bump, the cross 
section is enhanced by about 20\% over the standard-model level.

\begin{figure}
\centerline{\BoxedEPSF{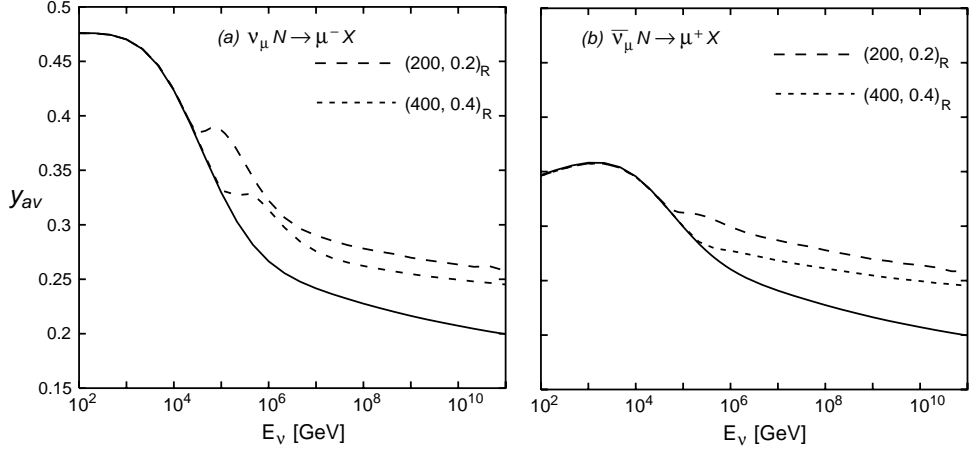 scaled 1350}} \caption[cc_y]{Mean 
inelasticity parameter $y_{\mathrm{av}}$ for (a) $\nu_{\mu}N$ 
and (b) $\bar{\nu}_{\mu}N$ charged-current interactions.  The 
solid lines show the predictions of the standard model.  The dashed 
(short-dashed) curves include the contributions of a right-handed 
squark, $\tilde{d}^{k}_{\mathrm{R}}$, with mass 
$\widetilde{m}=200\:(400)\gevcc$ and \rpv\ coupling 
$\lambda^{\prime}_{21k}= 0.2\:(0.4)$.} \label{fig:CC_y}
\end{figure}

The squark contributions modify the angular distributions as well.
In Figure \ref{fig:CC_y}(a), we plot the mean inelasticity 
$y_{\mathrm{av}} \equiv (1/\sigma)\int_{0}^{1}dy\,y(d\sigma/dy)$ as a function 
of $E_\nu$. The isotropic distribution of the squark decay products 
enhances $y_{\mathrm{av}}$ at the resonance bump and beyond.

\begin{figure}             
        \centerline{\BoxedEPSF{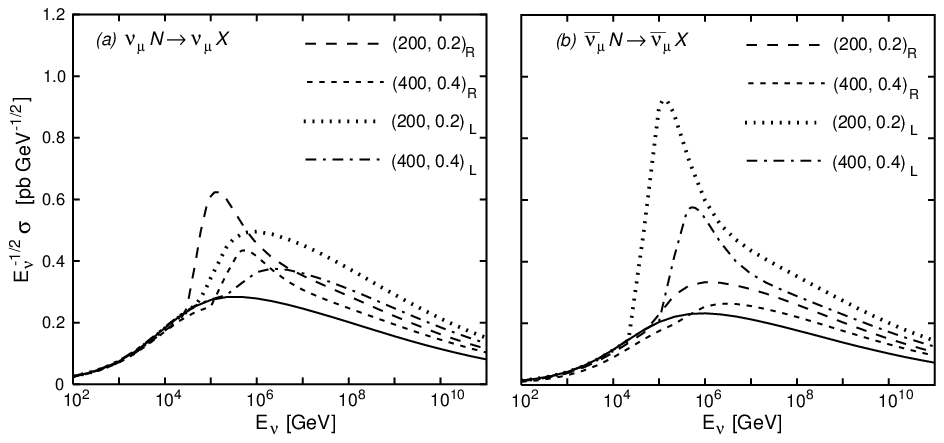 scaled 1350}} 
        \caption[nc]{Neutral-current cross sections for (a) 
        $\nu_{\mu}N$ and (b) $\bar{\nu}_{\mu}N$ interactions.  The 
        solid lines show the predictions of the standard model.  The dashed 
        (short-dashed) curves include the contributions of a right-handed 
        squark, $\tilde{d}^{k}_{\mathrm{R}}$, with mass 
        $\widetilde{m}=200\:(400)\gevcc$ and \rpv\ coupling 
        $\lambda^{\prime}_{21k}= 0.2\:(0.4)$.  The dotted (dot-dashed) curves 
        include the contributions of a left-handed squark, 
        $\tilde{d}^{k}_{\mathrm{L}}$, with mass 
        $\widetilde{m}=200\:(400)\gevcc$ and \rpv\ coupling 
        $\lambda^{\prime}_{2k1}= 0.2\:(0.4)$.} \label{fig:NC}
\end{figure}

The right-handed squark $\tilde{d}^{k}_{\mathrm{R}}$ has a similar 
influence on the neutral-current reaction $\nu_{\mu}N \rightarrow 
\nu_{\mu}+\hbox{anything}$.  There is an $s$-channel contribution 
from the formation process $\nu_{\mu}d_{\mathrm{L}}\rightarrow 
\tilde{d}^{k}_{\mathrm{R}} \rightarrow \nu_{\mu}d_{\mathrm{L}}$, which 
involves valence quarks, and a $u$-channel contribution to the 
reaction $\nu_{\mu}\bar{d} \rightarrow \bar{d}\nu_{\mu}$, which 
involves sea quarks.  Again, the resonance bumps shown in Figure 
\ref{fig:NC}(a) significantly enhance the standard-model 
cross section: the neutral-current cross section is increased by as 
much as 120\% and 55\% in our two examples.  Far above the resonance 
bumps, the increase is about 40\%.
The $\tilde{d}^{k}_{\mathrm{R}}$ contribution is more visible than in 
the charged-current case of Figure \ref{fig:CC}(a) because 
$\sigma_{\mathrm{NC}}/\sigma_{\mathrm{CC}} \approx 0.4$ in the standard 
model, whereas the \rpv\ cross sections are the same for 
neutral-current and charged-current interactions.

Left-handed squarks can contribute to the neutral-current reaction.  
An $s$-channel contribution arises from $\tilde{d}^{kc}_{\mathrm{L}}$ 
formation in the reaction $\nu_{\mu} \bar{d}_{\mathrm{R}} \rightarrow 
\tilde{d}^{kc}_{\mathrm{L}} \rightarrow \nu_{\mu} \bar{d}_{\mathrm{R}}$ 
involving the light-quark sea of the nucleon, while $u$-channel 
$\tilde{d}^{k}_{\mathrm{L}}$-exchange drives the reaction $\nu_{\mu} 
d_{\mathrm{R}}\rightarrow d_{\mathrm{R}}\nu_{\mu}$ involving valence 
quarks as well.  As we did for $\tilde{d}^{k}_{\mathrm{R}}$, we 
consider squark masses $\widetilde{m}^{k}_{\mathrm{L}} = 200\hbox{ and 
}400\gevcc$, and \rpv\ Yukawa couplings $\lambda^{\prime}_{2k1}= 
0.2\hbox{ and }0.4$.  In distinction to the 
$\tilde{d}^{k}_{\mathrm{R}}$, the $\tilde{d}^{kc}_{\mathrm{L}}$ decays 
only into $\nu_{\mu}\bar{d}$ through \rpv\ interactions; its total 
width $\Gamma(\tilde{d}^{k}_{\mathrm{L}}) = \frac{1}{2}
\Gamma(\tilde{d}^{k}_{\mathrm{R}})$.  The influence of 
$\tilde{d}^{k}_{\mathrm{L}}$ on the $\nu N$ neutral-current cross 
section is shown in Figure \ref{fig:NC}(a).  The unit 
branching fraction into the neutral-current mode means that the effect 
is pronounced, even though the $s$-channel reaction involves only sea 
quarks.  Because the sea is softer than the valence, the influence of 
$\tilde{d}^{k}_{\mathrm{L}}$ is deferred to energies well above the 
nominal threshold.  The 
maximum enhancement over the standard model is about 80\% (50\%) for 
$\widetilde{m}^{k}_{\mathrm{L}}=200\:(400)\gevcc$.  This enhancement 
persists to the highest energies we consider.

\begin{figure}             
        \centerline{\BoxedEPSF{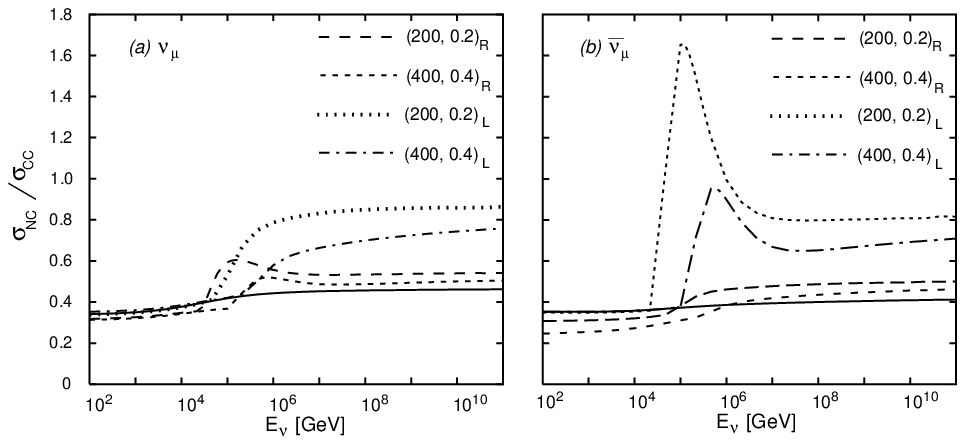 scaled 1350}} 
        \caption[ratio]{Neutral-current to charged-current ratios for 
        (a) $\nu_{\mu}N$ and (b) $\bar{\nu}_{\mu}N$ 
        interactions.  The solid lines show the predictions of the standard 
        model.  The dashed (short-dashed) curves include the contributions of 
        a right-handed squark, $\tilde{d}^{k}_{\mathrm{R}}$, with mass 
        $\widetilde{m}=200\:(400)\gevcc$ and \rpv\ coupling 
        $\lambda^{\prime}_{21k}= 0.2\:(0.4)$.  The dotted (dot-dashed) curves 
        include the contributions of a left-handed squark, 
        $\tilde{d}^{k}_{\mathrm{L}}$, with mass 
        $\widetilde{m}=200\:(400)\gevcc$ and \rpv\ coupling 
        $\lambda^{\prime}_{2k1}= 0.2\:(0.4)$.} \label{fig:Ratio}
\end{figure}

The neutral-current to charged-current ratio is an interesting 
diagnostic for the new physics represented by \rpv\ couplings.  In 
Figure \ref{fig:Ratio}(a) we compare the ratio 
$\sigma_{\mathrm{NC}}/\sigma_{\mathrm{CC}}$ expected in the standard 
model with our four \rpv\ examples.  The most prominent feature is the 
rapid rise of $\sigma_{\mathrm{NC}}/\sigma_{\mathrm{CC}}$ at the onset 
of $\tilde{d}^{k}_{\mathrm{L}}$ formation; the neutral-current to 
charged-current ratio is roughly doubled for squarks of 200 or 
$400\gevcc$
.

The ability to measure the neutral-current to charged-current ratio 
has not, until now, been conceived as a strength of neutrino 
telescopes.  Our study of \rpar--violating interactions shows that it 
would add markedly to the sensitivity that neutrino observatories will 
have to new physics.

\textit{$\bar{\nu}_{\mu}N$ interactions.} The charged-current reaction 
$\bar{\nu}_{\mu}N \rightarrow \mu^{+}+\hbox{anything}$ can receive 
contributions from the $s$-channel formation reaction 
$\bar{\nu}_{\mu}\bar{d}_{\mathrm{L}}\rightarrow \tilde{d}^{kc}_{\mathrm{R}} 
\rightarrow \mu^{+}\bar{u}_{\mathrm{L}}$, which involves sea quarks, and from 
$u$-channel exchange of $\tilde{d}^{kc}_{\mathrm{R}}$ in the reaction 
$\bar{\nu}_{\mu}u_{\mathrm{L}}\rightarrow d_{\mathrm{L}}\mu^{+}$, which involves valence 
quarks.  For the cases we consider, illustrated in Figure 
\ref{fig:CC}(b), the enhancement over the standard-model cross section 
is no more than 20\% (15\%) for 
$\widetilde{m}^{k}_{\mathrm{R}}=200\:(400)\gevcc$.

The right-handed squark $\tilde{d}^{kc}_{\mathrm{R}}$ has a similar 
influence on the neutral-current reaction $\bar{\nu}_{\mu}N 
\rightarrow \bar{\nu}_{\mu}+ \hbox{anything}$.  There is an 
$s$-channel contribution from the formation process 
$\bar{\nu}_{\mu}\bar{d}_{\mathrm{L}} \rightarrow \tilde{d}^{kc}_{\mathrm{R}} 
\rightarrow \bar{\nu}_{\mu}\bar{d}_{\mathrm{L}}$, which involves only sea quarks, 
and a $u$-channel contribution to the reaction $\bar{\nu}_{\mu}d _{\mathrm{L}} 
\rightarrow d _{\mathrm{L}} \bar{\nu}_{\mu}$, which involves valence quarks as well.  
The changes to the standard-model cross sections shown in Figure 
\ref{fig:NC}(b) are relatively more important than in the 
charged-current case, because the standard-model neutral-current cross 
section is smaller.  The maximum enhancements are 50\% (30\%) for 
$\widetilde{m}^{k}_{\mathrm{R}}=200\:(400)\gevcc$.  We see in Figure 
\ref{fig:Ratio}(b) that the right-handed squark does not lead to a 
dramatic effect in the neutral-current to charged-current ratio.

Left-handed squarks can alter the neutral-current cross section 
significantly.  An $s$-channel contribution arises from the formation 
reaction $\bar{\nu}_{\mu}d_{\mathrm{R}} \rightarrow 
\tilde{d}^{k}_{\mathrm{L}} \rightarrow 
\bar{\nu}_{\mu}d_{\mathrm{R}}$ involving valence quarks, and $u$-channel 
$\tilde{d}^{k}_{\mathrm{L}}$ exchange drives the reaction
$\bar{\nu}_{\mu} \bar{d}_{\mathrm{R}} \rightarrow 
\bar{d}_{\mathrm{R}}\bar{\nu}_{\mu}$,   involving 
sea quarks only.  The influence of $\tilde{d}^{k}_{\mathrm{L}}$  on the 
$\bar{\nu}_{\mu}N$ neutral-current reaction is shown in Figure 
\ref{fig:NC}(b).  For the cases we consider, \rpv\ contributions can 
be 3 times (2 times) the standard-model cross section, for
$\widetilde{m}^{k}_{\mathrm{L}}=200\:(400)\gevcc$.  As a consequence, 
the effect of a left-handed squark on the neutral-current to 
charged-current ratio is extremely pronounced: we see in Figure 
\ref{fig:Ratio}(b) that $\sigma_{\mathrm{NC}}/\sigma_{\mathrm{CC}}$
is quadrupled (doubled) for our two examples.  Far above threshold, 
$\sigma_{\mathrm{NC}}$ is increased by 95\% (75\%), and the 
neutral-current to charged-current ratio remains twice ($1\frac{3}{4}$ 
times) the standard-model value.

\begin{figure}          
        \centerline{\BoxedEPSF{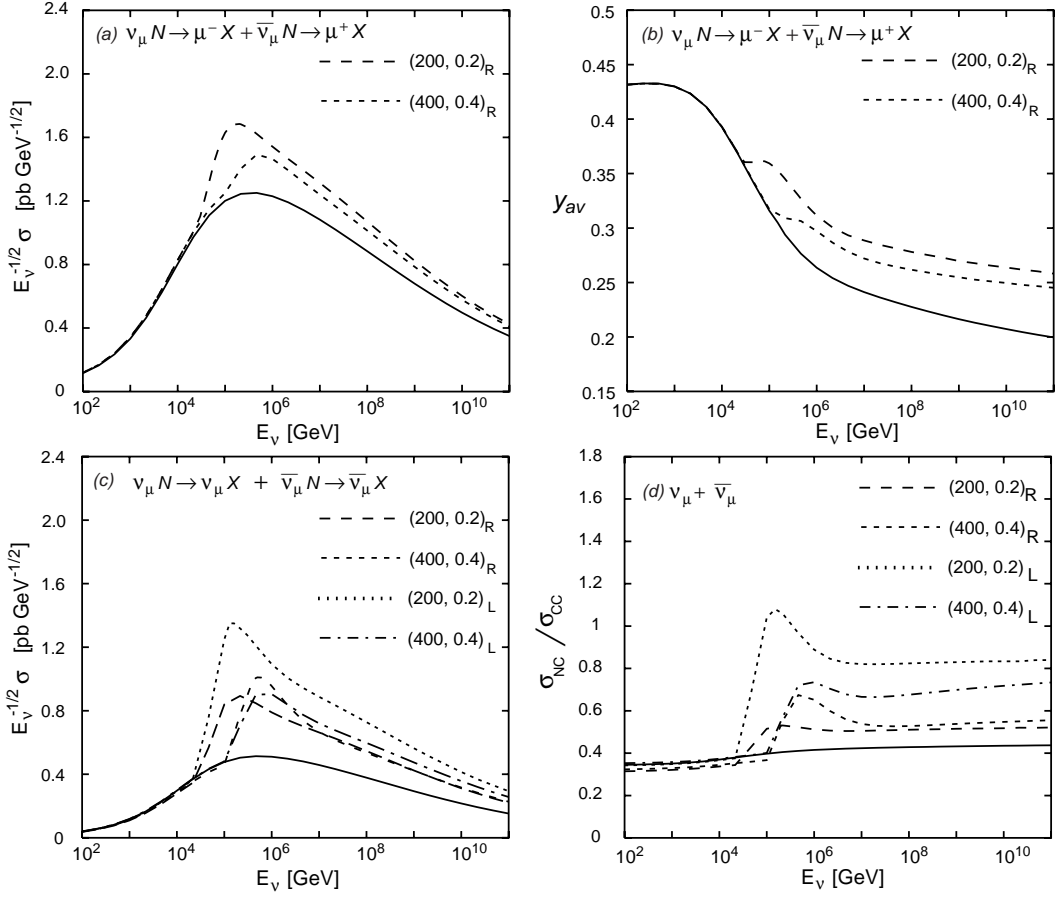 scaled 1350}} 
        \caption[ratio]{Observables for $\nu N +\bar{\nu}N$ interactions.  (a) 
        charged-current cross section; (b) mean inelasticity $y_{\mathrm{av}}$ 
        for charged-current interactions; (c) neutral-current cross section; 
        and (d) neutral-current to charged-current ratio.  The solid lines 
        show the predictions of the standard model.  The dashed (short-dashed) 
        curves include the contributions of a right-handed squark, 
        $\tilde{d}^{k}_{\mathrm{R}}$, with mass 
        $\widetilde{m}^{k}_{\mathrm{R}}=200\:(400)\gevcc$ and \rpv\ coupling 
        $\lambda^{\prime}_{21k}= 0.2\:(0.4)$.  The dotted (dot-dashed) curves 
        include the contributions of a left-handed squark, 
        $\tilde{d}^{k}_{\mathrm{L}}$, with mass 
        $\widetilde{m}^{k}_{\mathrm{L}}=200\:(400)\gevcc$ and \rpv\ coupling 
        $\lambda^{\prime}_{2k1}= 0.2\:(0.4)$.} \label{fig:averages}
\end{figure}

\textit{Consequences for neutrino observatories.}
Until now, we have considered $\nu_\mu N$ and $\bar{\nu}_\mu N$ 
interactions separately.  Neutrino telescopes will not distinguish 
between events induced by neutrinos and antineutrinos.  Thus we should 
rather consider the sum of the $\nu_\mu N$ and $\bar{\nu}_\mu N$ cross 
sections.  The four panels in Figure \ref{fig:averages} depict 
$\sigma_{\mathrm{CC}}$, $y_{\mathrm{av}}$ for charged-current 
interactions, $\sigma_{\mathrm{NC}}$, and 
$\sigma_{\mathrm{NC}}/\sigma_{\mathrm{CC}}$ for summed $\nu_{\mu}N + 
\bar{\nu}_{\mu}N$ interactions.  The significance of the resonance 
threshold is reduced, but the deviations from the standard-model 
expectations remain significant.  The 
increases in the mean inelasticity and the neutral-current to 
charged-current ratio seem especially promising signals for 
unconventional interactions, because they do not require a knowledge 
of the incident neutrino flux.

\section{Neutrino interactions on electron targets}
Because of the electron's small mass, neutrino-electron interactions 
in matter can generally be neglected with respect to neutrino-nucleon 
interactions \cite{nue}.  There is one exceptional case in the 
standard model: resonant formation of the intermediate boson in 
$\bar{\nu}_{e}e \rightarrow W^{-}$ interactions at $6.3\pev$.  The 
resonant cross section is larger than the $\nu N$ cross section at any 
energy up to $10^{21}\ev$.  This singular case makes it important to 
bear the neutrino-electron cross sections in mind when assessing the 
capabilities of neutrino telescopes.  We now consider similar singular 
cases that can arise through \rpar--violating interactions.  Once we 
allow $LLE$ couplings, a wide variety of new production and decay 
channels open up.  Because the \rpv\ couplings are constrained to be 
small, only channels that involve resonant sparticle production can 
display sizeable effects.

{\it $\nu e$ interactions.}  It follows from the interaction 
Lagrangian \eqn{eq:LLE} that a left-handed slepton 
can only give rise to additional $u$-channel diagrams generated by 
the operator $L_{i}L_{j}\bar{E}_{k}$, with $i \neq j$ from SU(2) 
invariance.  \rpar--violating contributions to the reactions 
$(\nu_{\mu},\nu_{e})e \rightarrow (\nu_{\mu},\nu_{e})e$ are mediated 
by $\tilde{\tau}_{\mathrm{L}}$ or 
$(\tilde{e}_{\mathrm{L}},\tilde{\mu}_{\mathrm{L}})$.  These lead to 
unobservably small modifications to the standard-model cross section.

In contrast, a right-handed slepton can appear in an $s$-channel 
diagram, and may thus be produced as a resonance.  Because of the 
SU(2) invariance of the $L_{i}L_{j}\bar{E}_{k}$ operator, the 
options are $(\nu_{\mu},\nu_{\tau})e \rightarrow 
\tilde{e}^{k}_{\mathrm{R}}$.  For the astrophysically interesting case 
of incident $\nu_{\mu}$, and considering only one \rpv\ operator at a 
time, the two final states $\nu_{\mu}e$ and $\nu_{e}\mu$ occur with 
equal probability in the decay of $\tilde{e}^{k}_{\mathrm{R}}$.  Within 
the standard model, $\nu_{\mu}e \rightarrow \nu_{\mu}e$ elastic 
scattering proceeds through a $t$-channel $Z$-exchange graph.  At low 
energies, the cross section grows with neutrino energy, but reaches a 
plateau at $10\pb$ around $E_{\nu}=10^{17}\ev$.  The charge-exchange 
reaction $\nu_{\mu}e \rightarrow \mu \nu_{e}$ proceeds by 
$W$-exchange, which leads to a cross section about an order of 
magnitude larger.  In either channel, the magnitude of the resonant 
cross section is governed by the strength of the \rpv\ coupling 
$\lambda_{12k}$ and the slepton mass.  Before discussing the 
observability of the $\nu_{\mu}e_{\mathrm{L}} \rightarrow 
\tilde{e}^{k}_{\mathrm{R}} \rightarrow (\nu_{\mu}e_{\mathrm{L}}, 
\nu_{e}\mu_{\mathrm{L}})$ signals, let us explore the signals that 
arise in $\bar{\nu}_{e}e$ and $\bar{\nu}_{\mu}e$ interactions.

{\it $\bar{\nu}_{e}e$ and $\bar{\nu_{\mu}}e$ interactions.} Resonant 
slepton production can occur in the reactions 
$\bar{\nu}_{e}e_{\mathrm{R}} \rightarrow 
\tilde{\tau}_{\mathrm{L}}\hbox{ or }\tilde{\mu}_{\mathrm{L}}$ and 
$\bar{\nu}_{\mu}e_{\mathrm{R}} \rightarrow 
\tilde{\tau}_{\mathrm{L}}\hbox{ or }\tilde{e}_{\mathrm{L}}$.  If we 
suppose that only one \rpv\ coupling is nonzero, then the left-handed 
slepton cannot mediate flavor-changing processes such as 
$\bar{\nu}_{e}e \rightarrow \bar{\nu}_{\mu}\mu$.

\begin{figure}            
        \centerline{\BoxedEPSF{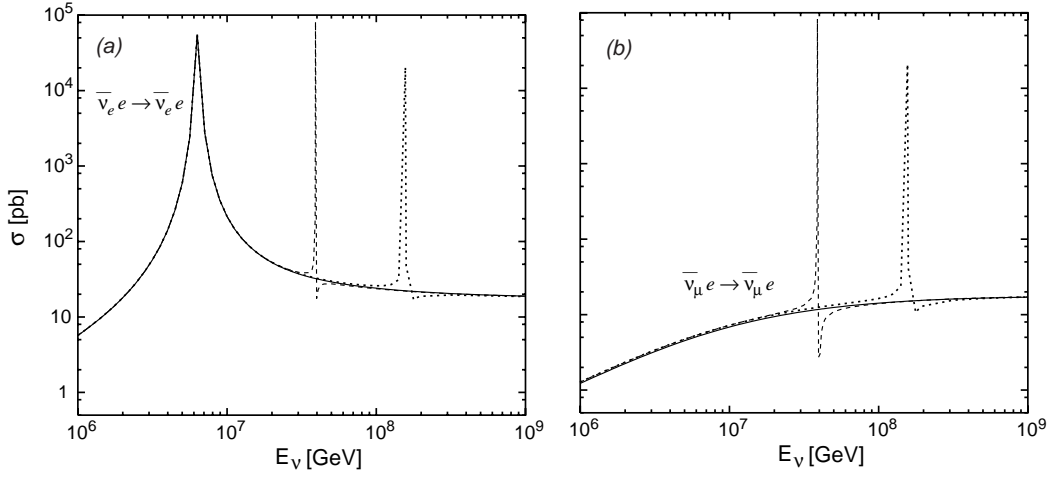 scaled 850}} \caption[sm_e]{(a) 
        Cross section for $\bar{\nu}_{e}e$ elastic scattering in the standard 
        model (solid line), showing the $W^{-}$ resonance at $6.3\pev$, and 
        for resonant production of a 200-GeV$\!/\!c^{2}$ (short dashes) or 
        400-GeV$\!/\!c^{2}$ (dotted line) slepton through an \rpv\ 
        coupling.  (b) $\bar{\nu}_{\mu}e \rightarrow \bar{\nu}_{\mu}e$ cross 
        sections for resonant production of a 200-GeV$\!/\!c^{2}$ (short 
        dashes) or 400-GeV$\!/\!c^{2}$ (dotted line) slepton through 
        an \rpv\ coupling, compared with the featureless standard-model 
        prediction (solid line).  The \rpv\ couplings are $\lambda = 
        (0.1, \;0.2)$ for $m_{\tilde{e}^{k}_{\mathrm{L}}} = (200,\;400)\gevcc$.}
        \label{fig:e_nu_sm}                                                   
\end{figure}

In Figure \ref{fig:e_nu_sm} we present the cross sections for (a) 
$\bar{\nu}_{e}e$ and (b) $\bar{\nu}_{\mu}e$ elastic scattering.  
Within the standard model, $\bar{\nu}_{e}e$ elastic scattering is 
driven by $s$-channel $W^{-}$ formation and $t$-channel $Z^{0}$ 
exchange; the behavior of $\bar{\nu}_{\mu}e$ elastic scattering is 
very similar to that of the reaction $\nu_{\mu}e \rightarrow 
\nu_{\mu}e$ described above.  A left-handed slepton with a single 
\rpv\ coupling may occur as an $s$-channel resonance in either 
$\bar{\nu}_{e}e$ elastic scattering or $\bar{\nu}_{\mu}e$ elastic 
scattering, but not both.  If \rpar--\textit{conserving} decays are 
absent, as assumed in Figure \ref{fig:e_nu_sm}, the slepton resonance 
is very narrow and impressively tall.  The peak cross section is
\begin{equation}
	\sigma_{\mathrm{peak}}(\bar{\nu}_{i}e \rightarrow \bar{\nu}_{i}e) \approx 
	\frac{8\pi}{m_{\tilde{e}^{k}_{\mathrm{L}}}^{2}}\cdot \left(
	\frac{\Gamma(\tilde{e}^{k}_{\mathrm{L}}\rightarrow \bar{\nu}_{i}e)}
	{\Gamma_{\mathrm{tot}}(\tilde{e}^{k}_{\mathrm{L}})}\right)^{2} \; ,
	\label{eq:peak}
\end{equation}
where the final factor is unity, by assumption.  The total width of 
the slepton resonance is $\Gamma_{\mathrm{tot}}(\tilde{e}^{k}_{\mathrm{L}})= 
(\lambda^{2}/16\pi)m_{\tilde{e}^{k}_{\mathrm{L}}}$.
For the case of $\nu_{\mu}e \rightarrow 
\tilde{e}^{k}_{\mathrm{R}}$ scattering discussed above, the peak cross 
section is similar in form.  In that case, however, 
$\tilde{e}^{k}_{\mathrm{R}}$ decays with equal probability to $\nu_{\mu}e$ or 
$\nu_{e}\mu$.  The peak cross section for $\nu_{\mu}e \rightarrow 
\nu_{\mu}e$ or $\nu_{\mu}e \rightarrow \nu_{e}\mu$ is accordingly 
rescaled by a factor 1/4, and the total width is doubled.

\textit{Detecting the slepton resonance.} The modest flux of UHE 
neutrinos and the factor-of-two energy resolution anticipated for 
neutrino observatories mean that it will be difficult to resolve such 
a narrow structure from the standard-model background.  One 
distinguishing characteristic is that the slepton resonance will only 
be produced in downward-going interactions.  In water-equivalent 
units, the interaction length is
\begin{equation}
	L_{\mathrm{int}}^{(e)} = \frac{1}{\sigma(E_{\nu})(10/18)
	N_{\mathrm{A}}}\; ,
	\label{eq:lint}
\end{equation}
where $N_{\mathrm{A}}=6.022\times 10^{23}\hbox{ mol}^{-1} = 
6.022\times 10^{23}\cm^{-3}$ (water equivalent) is Avogadro's number 
and $(10/18)N_{\mathrm{A}}$ is the number of electrons in a mole of 
water.  At the peak of a 200- (400-)$\gevcc$ slepton resonance 
produced in $\bar{\nu}e$ interactions, the 
interaction length is 122 (488)$\kmwe$.  The resonance is effectively 
extinguished for neutrinos that traverse the Earth, whose diameter 
is $1.1\times 10^{5}\kmwe$.  Similar remarks apply to the 
$\nu_{\mu}e_{\mathrm{L}} \rightarrow \tilde{e}^{k}_{\mathrm{R}} \rightarrow 
(\nu_{\mu}e_{\mathrm{L}}, \nu_{e}\mu_{\mathrm{L}})$ cases.

A more promising strategy to observe a slepton resonance is to 
consider final states that would clearly stand out above the 
background.  One such possibility arises if both $LLE$ and $LQD$ 
operators exist.  Then the slepton could decay into two jets, for 
which the background from $\bar{\nu}_{e}e \rightarrow W^{-} 
\rightarrow q\bar{q}$ is negligible at high energies.  
Neutral-current $\nu N \rightarrow \nu+\hbox{anything}\;$ interactions 
might constitute an additional---and large---background.

A much more interesting possibility arises if neutralinos are relatively light.
Now the slepton may also decay into the corresponding 
lepton and a light neutralino, which decays in turn into leptons 
and neutrinos.  Reactions that produce final-state $\tau$-leptons are 
of special interest.  The decay length of a 1-PeV $\tau$ is about 50 
meters, so the production and subsequent decay of a $\tau$ at ultrahigh 
energies will give rise to a characteristic ``double-bang'' signature 
in a water or ice \v{C}erenkov detector.  We consider
\dkp{\nu_{\mu}e^{-}_{\mathrm{L}} \rightarrow 
\tilde{\tau}^{-}_{\mathrm{R}}\rightarrow}{\tau^{-}}{\tilde{\chi}^{0}}
{\tau^{+}_{\mathrm{R}}\nu_{e}\mu^{-}_{\mathrm{L}}\hbox{ or }
\tau^{+}_{\mathrm{R}}\nu_{\mu}e^{-}_{\mathrm{L}}}
and
\dkp{\bar{\nu}e^{-}_{\mathrm{R}} \rightarrow 
\tilde{\tau}^{-}_{\mathrm{L}}\rightarrow}{\tau^{-}}{\tilde{\chi}^{0}}
{\tau^{+}_{\mathrm{L}}\bar{\nu} e^{-}_{\mathrm{R}}\;,}
where $\bar{\nu}$ refers to $\bar{\nu}_{e}$ or $\bar{\nu}_{\mu}$.
Close to the resonance (\ie,  where the $t$- and $u$- channel 
exchanges can be neglected in comparison to the $s$-channel pole), 
the cross section for $\tau  \tilde{\chi}^{0}$ production can be 
approximated by a Breit-Wigner formula,
\begin{eqnarray}
	\sigma & = & \frac{8 \pi s}{m_{\tilde{\tau}}^2}
       \;
   \frac{\Gamma (\tilde{\tau} \rightarrow e^- \nu) 
         \: 
         \Gamma (\tilde{\tau} \rightarrow \tau^- \tilde{\chi}_{0})
        }
        {   (s - m_{\tilde{\tau}}^2)^2 
          + m_{\tilde{\tau}}^2 \Gamma_{\rm total}^2 
        }
        \;
     \left[ \frac{ s - m_{\tilde{\chi}^0 }^2   }
		 { m_{\tilde{\tau}}^2 - m_{\tilde{\chi}^0 }^2   }
     \right]^2
	\nonumber\\
	 & \rightarrow & \frac{8\pi}{m_{\tilde{\tau}}^{2}}
	 B(\tilde{\tau}\rightarrow \nu e) B(\tilde{\tau}\rightarrow 
	 \tau\tilde{\chi}^{0})\;\hbox{, as $s\rightarrow m_{\tilde{\tau}}^{2}$.}
	\label{BWneut}
\end{eqnarray}
The signal cross section can thus be deduced from 
Figure \ref{fig:e_nu_sm}(b), scaled by the appropriate branching 
fractions.  The standard-model background for $\tau$ production 
arises only from the reaction $\bar{\nu}_{e}e \rightarrow W^{-} 
\rightarrow \bar{\nu}_{\tau}\tau$; it drops below a few$\pb$ at about 
$10^{17}\ev$.  To good approximation, the production of taus through a 
slepton resonance with a mass of $200\gevcc$ or greater is background-free.

\begin{figure}
 \centerline{\BoxedEPSF{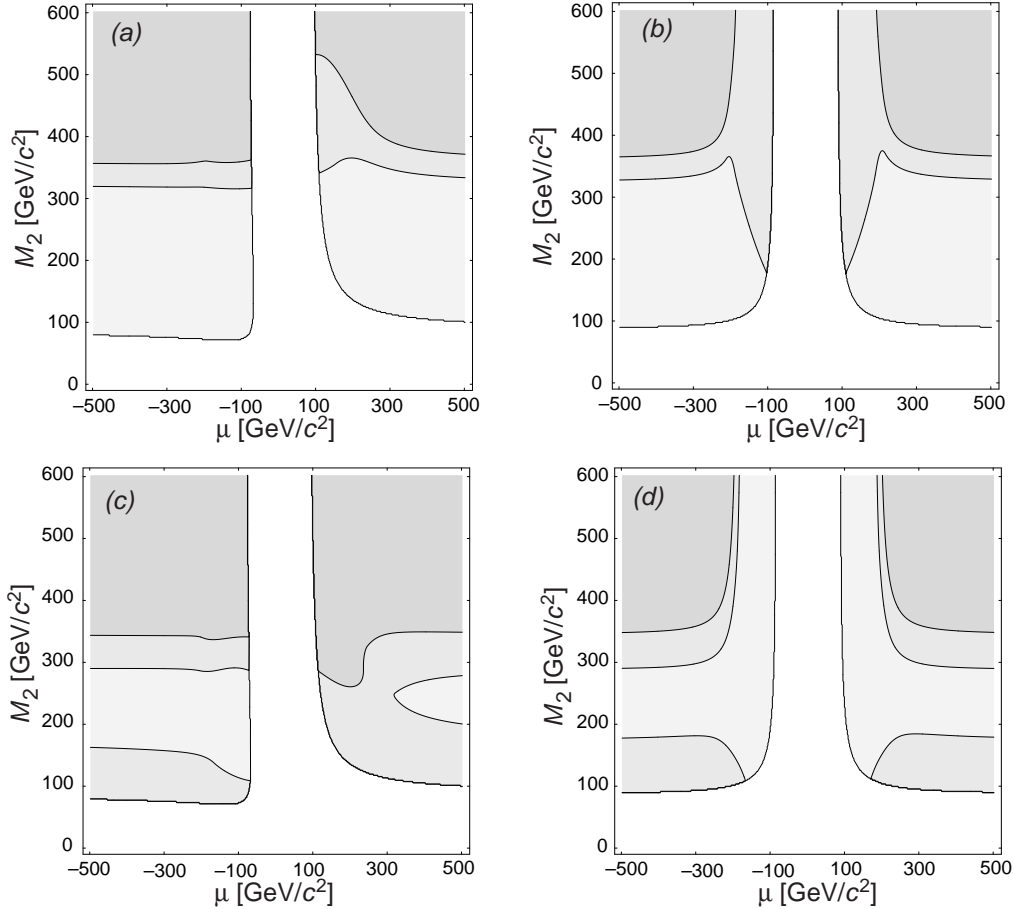 scaled 800}} 
\caption[cont]{Branching ratios for $\tilde{\tau}$ decays to 
$\tau+\hbox{neutralinos}$ for $m_{\tilde{\tau}_{\mathrm{R,L}}}=200\gevcc$ and 
$\lambda = 0.1$, as functions of parameters of the minimal 
supergravity model.  The four cases correspond to: (a) 
$\tilde{\tau}_{\mathrm{R}}$ production with $\tan\beta = $2, (b) 
$\tilde{\tau}_{\mathrm{R}}$ production with $\tan\beta = 60$, (c) 
$\tilde{\tau}_{\mathrm{L}}$ production with $\tan\beta = 2$, and (d) 
$\tilde{\tau}_{\mathrm{L}}$ production with $\tan\beta = 60$.  The 
shaded regions denote branching fractions $0-0.3$ (dark), $0.3-0.6$ 
(medium), $0.6-1.0$ (light).  The U(1) gaugino mass $M_1$ is 
determined in terms of the SU(2) gaugino mass $M_2$ by the 
unification relation $M_1 = (5/3)M_2\tan^2\theta_W$.  We exclude 
regions that correspond to chargino masses below the LEP2 bound of 
$85\gevcc$.}
         \label{fig:Contours}                                                        
\end{figure}

In Figure \ref{fig:Contours} we plot branching-fraction contours for 
the decays $\tilde{\tau}_{\mathrm{L,R}} \rightarrow \tilde{\chi}^{0} 
\tau_{\mathrm{L,R}}$ in the $M_{2}-\mu$ plane for small and large 
values of $\tan\beta$, the ratio of the vacuum expectation value of the 
Higgs boson that couples to up quarks to that of the Higgs boson that 
couples to down quarks.  Here $M_{2}$ is the SU(2) gaugino mass parameter 
and $\mu$ is the Higgsino mass parameter.  The \rpar--conserving decay 
mode competes with, and even dominates, the \rpv\ decay mode(s) for a 
wide range of supersymmetric parameters.  For 
$\tilde{\tau}_{\mathrm{R}}$, the \rpv\ decays are divided equally 
between the $\nu_{e}\mu^{-}$ and $\nu_{\mu}e^{-}$ modes, but the 
$R$-conserving decay into charginos is very suppressed.  At small and 
moderate values of $M_{2}$ and $\mu$, the tau-neutralino decay 
dominates and the total width of $\tilde{\tau}_{\mathrm{R}}$ can be as 
large as $0.9\gev$.  For $\tilde{\tau}_{\mathrm{L}}$, the only \rpv\ 
decay is into $\nu_{e}\mu^{-}$, but the $R$-conserving decays can 
include both the $\tau\tilde{\chi}^{0}$ mode and the 
$\nu_{\tau}\tilde{\chi}^{-}$ mode.  The best bet for unconventional 
tau production thus seems to be through the production of 
$\tilde{\tau}_{\mathrm{R}}$ in $\nu_{\mu}e$ collisions.

Because we know of no conventional astrophysical sources of tau 
neutrinos, the observation of an excess of energetic taus in a neutrino telescope 
has been suggested as a signature for neutrino oscillations 
\cite{pandl}.  We learn from the above discussion that if a massive 
slepton is produced through \rpv\ interactions, it may well be a 
source of energetic taus that represents a different sort of physics 
beyond the standard model.

\section{Conclusion}
Supersymmetric theories with \rpar--violation are an interesting 
generalization of the usual physics beyond the standard model.  
Relaxing the assumption that \rpar\ is conserved alters the 
phenomenology of supersymmetry in a crucial way.  In accelerator 
searches for supersymmetric particles, the missing energy signature is 
removed because the lightest supersymmetric particle can decay into 
ordinary particles, decay chains are modified, and superpartners can 
be produced singly.  We have shown that \rpar--violating processes can 
also have significant effects on ultrahigh-energy neutrino cross 
sections, if the \rpv\ couplings are as large as allowed by current 
bounds.

We have considered the influence of squarks with \rpv\ couplings on 
the cross sections for UHE $\nu_{\mu}N$ and $\bar{\nu}_{\mu}N$ 
interactions, and the effect of sleptons with \rpv\ couplings on 
$\nu_{\mu}e$, $\bar{\nu}_{e}e$ and $\bar{\nu}_{\mu}e$ interactions.  
For the case of neutrino-nucleon collisions, we have found that the 
neutral-current to charged-current ratio can be a very useful 
diagnostic for new physics.  The ability to measure neutral-current 
interactions and characterize their energy should receive increased 
attention in design studies for neutrino telescopes.

A remarkable sign for \rpar--violating effects in neutrino-electron 
scattering would be the formation of slepton resonances in narrow 
energy bands.  The large cross sections at resonance might be observed 
through the extinction of neutrinos passing through the Earth near the 
resonance energy or by the anomalous production of tau leptons in 
downward-going events only.
We have explored scenarios that lead to the production of energetic 
taus through \rpar--violating neutrino-electron interactions.  If a 
$\tilde{\tau}$ has both \rpar--conserving and 
\rpar--violating decays, we would expect resonant production of 
$\tau\tilde{\chi}^{0}$, with a subsequent decay of the LSP into 
$\tau\ell\nu$.  Alternatively, if more than one \rpv\ 
coupling is appreciable, we could expect resonant 
$\tilde{e}^{k}\rightarrow\tau\nu$ production. Either could 
provide an explanation for $\tau$ events that does not invoke neutrino 
oscillations.  The key to separating the two \rpv\ effects lies in observing 
the decay products of the $\tilde \chi^0$ by searching for associated 
leptons in $\tau$-bearing events.

It is plain that the observation of the unconventional effects we have 
examined here is some time away, even if sizeable \rpv\ couplings do 
occur in Nature.  Our intent is to widen the scientific horizons of 
neutrino observatories and to encourage imaginative thinking 
about the range of measurements they might make.

\section*{Acknowledgements}
The work of SL is funded by a Marie Curie fellowship 
(TMR-ERBFMBICT 959565).  MC, SL, and CQ acknowledge the hospitality of 
the Aspen Center for Physics, and DC acknowledges the hospitality of 
the Fermilab Theoretical Physics Department, where parts of this work 
were performed.  Fermilab is operated by Universities Research 
Association Inc.\ under Contract No.\ DE-AC02-76CH03000 with the 
United States Department of Energy.
\appendix
\section*{Cross section formulae}
It is straightforward to calculate the inclusive cross section for the 
reactions $\nu_{\mu}N \rightarrow \mu+\hbox{anything}$ and $\nu N 
\rightarrow \nu+\hbox{anything}$, where $N=\half(p+n)$ is an 
isoscalar nucleon, in the renormalization-group--improved parton 
model.  The differential cross section is written in terms of the 
Bjorken scaling variables $x=Q^{2}/2ME_{\nu}$ and $y=\nu/E_{\nu}$ 
as
\begin{equation}
        \frac{d^{2}\sigma}{dxdy} = \frac{ME_{\nu}}{16\pi}\:
          \sum_f 
          \left[ \left| a_f \right|^2 
                 + \left| b_f \right|^2 ( 1 - y)^2 
         \right] \; x f(x, Q^2) ,
        \label{dsdxy}
\end{equation}
where $-Q^{2}$ is the squared invariant momentum transfer between the 
incident and outgoing lepton, $\nu = E_{\nu}-E^{\prime}$ (with 
$E^{\prime}$ the energy of the outgoing lepton in the target frame) is 
the lepton energy loss, $M_{W}$ is the $W$-boson mass, and 
$G_{F}=1.16632 \times 10^{-5}\gev^{-2}$ is the Fermi constant.  Here 
$f(x, Q^2)$ represents the parton densities within the nucleon; the 
coefficients $a_f$ and $b_f$ for interactions on parton species $f$ 
depend on the particular process under consideration.

For charged-current interactions, the 
only nonzero coefficients are
\be
\barr{rcl} 
\dis a^{\mathrm{CC}}_{d_j} 
      & = & \dis 
           \frac{g^2}{Q^2 + M_W^2}
            - \sum_k 
                \frac{ \left| \lambda^{\prime}_{ijk} \right|^2}
                   { x s - \widetilde{m}^{k\;2}_{\mathrm{R}}
     + i\widetilde{m}^{k}_{\mathrm{R}}\Gamma(\tilde{d}^{k}_{\mathrm{R}})  
     }\;\; ,
    \\[2ex]
    
\dis b^{\mathrm{CC}}_{\bar u_j} 
      & = & \dis 
           \frac{g^2}{Q^2 + M_W^2}
            - \sum_k 
                \frac{ \left| \lambda^{\prime}_{ijk} \right|^2}
                   { Q^2 - x s - \widetilde{m}^{k\;2}_{\mathrm{R}} 
                   }\;\; ,
\earr 
\ee
where $s=2ME_{\nu}$ is the square of the c.m.\ energy, 
the square of the SU(2)$_{\mathrm{L}}$ gauge coupling is 
$g^{2}=8G_{F}M_{W}^{2}/\sqrt{2}$, and the $\lambda^{\prime}_{ijk}$ are 
the \rpv\ Yukawa couplings.

For neutral-current interactions, we have
\be
\barr{rcl} 
\dis a^{\mathrm{NC}}_{d_j} 
      & = & \dis 
           \frac{g^2}{2(1-x_{W})} \; \frac{L_d}{Q^2 + M_Z^2}
            - \sum_k 
                \frac{ \left| \lambda^{\prime}_{ijk} \right|^2}
                   { x s - \widetilde{m}^{k\;2}_{\mathrm{R}}
     + i \widetilde{m}^{k}_{\mathrm{R}}\Gamma(\tilde{d}^{k}_{\mathrm{R}})   }
    \\[2ex]
\dis b^{\mathrm{NC}}_{d_j} 
      & = & \dis 
           \frac{g^2}{2(1-x_{W})} \; \frac{R_d}{Q^2 + M_{Z}^2}
            + \sum_k 
                \frac{ \left| \lambda^{\prime}_{ijk} \right|^2}
                   { Q^2 - x s - \widetilde{m}^{k\;2}_{\mathrm{L}} }
    \\[2ex]
\dis a^{\mathrm{NC}}_{u_j} 
      & = & \dis 
           \frac{g^2}{2(1-x_{W})} \; \frac{L_u}{Q^2 + M_{Z}^2}
    \\[2ex]
\dis b^{\mathrm{NC}}_{u_j} 
      & = & \dis 
           \frac{g^2}{2(1-x_{W})} \; \frac{R_u}{Q^2 + M_{Z}^2}
\earr
\ee
where $M_{Z}$ is the $Z$-boson mass, $x_{W}\equiv \sin^{2}\theta_{W}$ 
is the weak mixing parameter, and the chiral couplings are $L_{u,d} = 
\pm 1 - 2x_{W}Q_{u,d}$ and $R_q = - 2x_{W}Q_q$, with $Q_{q}$ the 
electric charge of quark $q$.  For neutrino-antiquark interactions, 
the coefficients can be obtained by crossing symmetry, 
{\em viz.} $(a \leftrightarrow b, x s \leftrightarrow Q^2 - x s )$.

\newpage
\widetext

\newpage

\newpage
\newpage
\newpage
\narrowtext
\begin{table}[tb]
	\caption{Experimental constraints (at one or two standard deviations) on 
the \rpar--violating Yukawa couplings of interest, for the case of 
200-GeV$\!/\!c^{2}$ sfermions.  For arbitrary sfermion mass, multiply 
the limits by $(m_{\tilde{f}}/200\gevcc)$, except for 
$\lambda^{\prime}_{221}$.}
\begin{center}
	\begin{tabular}{cc}
			\rpv\ Coupling & Limited by  \\
			\hline
			$\lambda_{12k} < 0.1~(2\sigma)$ & charged-current universality  \\
			$\lambda_{131, 132, 231} < 0.12~(1\sigma)$ &  $\Gamma(\tau\rightarrow 
			e\nu\bar{\nu})/ \Gamma(\tau \rightarrow \mu\nu\bar{\nu})$  \\
			$\lambda_{133}< 0.006~(1\sigma)$ & $\nu_{e}$ Majorana mass \\[6pt]
			$\lambda^{\prime}_{21k} < 0.18~(1\sigma)$ & $\pi$ decay  \\
			$\lambda^{\prime}_{221} < 0.36~(1\sigma)$ & $D$ decay  \\
			$\lambda^{\prime}_{231} < 0.44~(2\sigma)$ & $\nu_{\mu}$ deep 
			inelastic scattering  \\
		\end{tabular}
	\end{center}  
\vspace*{24pt}
	\label{tab:lambdas}
\end{table}


\begin{thebibliography}{99}
\bibitem{e787} The first candidate $K^{+}\rightarrow 
\pi^{+}\nu\bar{\nu}$ event has been presented recently by S.  Adler, 
\etal\ (E787 Collaboration), \prl{79}{2204}{97}.  
Based on the observation of one event, the 90\% confidence level upper 
limit on the branching ratio is $B(K^{+}\rightarrow\pi^{+}\nu\bar{\nu}) < 
1.6 \times 10^{-9}$.
\bibitem{nutau}  Secondary $\bar{\nu}_{\tau}$ can be produced in the 
resonant $\bar{\nu}_{e}e \rightarrow W^{-} \rightarrow 
\bar{\nu}_{\tau}\tau$ reaction in the neighborhood of 
$E_{\bar{\nu}_{e}} = 6.3\pev$.  See \S 4 of Ref.\ \cite{GQRS}.
\end{thebibliography}
\end{document}